\documentclass[aps,prl,reprint,groupedaddress]{revtex4-1}
\usepackage{graphicx}
\usepackage[english]{babel}
\usepackage{graphicx}

\begin{document}
\title{Correlations of record events as a test for heavy-tailed distributions}
\author{J. Franke, G. Wergen, J. Krug}
\affiliation{Institute of Theoretical Physics, University of
  Cologne,Z\"ulpicher Strasse 77, 50937 K\"oln, Germany}
\begin{abstract}

A record is an entry in a time series that is larger or smaller than
all previous entries. If the time series consists of independent, identically distributed
random variables with a superimposed linear trend, record events are
positively (negatively) correlated when the tail of the distribution is
heavier (lighter) than exponential. Here we use these correlations to
detect heavy-tailed behavior in small sets of independent
random variables. The method consists of converting random subsets of 
the data into time series with a tunable linear drift and computing 
the resulting record correlations.

\end{abstract}

\date{\today}

\maketitle

Determining the probability distribution underlying a given data set or at
least its behavior for large argument is of pivotal importance for
predicting the behavior of the 
system: If the data is drawn from a distribution
with heavy tails, one needs to prepare for large events.
Of particular relevance is the case when the probability
density displays a power law decay, as this implies a drastic
enhancement of the probability of extreme events. This is one of the reasons
for the persistent interest in the observation and modeling of power law
distributions, which have been associated with critical, scale-invariant
behavior \cite{Sornette,Christensen} in diverse contexts ranging from
complex networks \cite{networks} to paleontology \cite{Newman1999}, 
foraging behavior of animals \cite{albatros}, citation distributions
\cite{r1998} and many more \cite{Clauset2009}. 

However, when trying to infer the tail behavior of the underlying distribution 
from a finite data set, one faces 
the problem that the number of entries of large absolute value is very
small. This implies that even though binning the
entries by magnitude and plotting them would yield
an approximate representation of the probability density, this process
becomes inconclusive in particular in the tail of the probability density.
Furthermore, in small data sets, extreme outliers can
strongly affect the results of methods like maximum likelihood
estimators such that leaving out even one of these extreme and
possibly spurious data points renders the outcome of the test
insignificant. A case in point is the problem of estimating the distribution
of fitness effects of beneficial mutations in evolution experiments, which
are expected on theoretical ground to conform to one of the universality
classes of extreme value theory (EVT) \cite{Joyce2008}. Because
beneficial mutations are rare, the corresponding data sets are
typically limited to a few dozen values, and the determination of the
tail behavior can be very challenging \cite{Rokyta2008,Miller2011}. 

In this Letter we present a method for detecting
heavy tails in empirical data that works reliably for small data
sets (on the order of a few dozen entries) and is robust with respect
to removal of extreme entries. The test is based on
the statistics of records of subsamples of the data set.
Similar to conventional record-based statistical tests
\cite{Foster1954,Glick1978,Gulati2003}, and in contrast to the bulk of
methods available in this field \cite{Clauset2009}, our approach is
non-parametric and, hence, does not require any hypothesis about the
underlying distribution. Rather than aiming at reliable estimates of
the parameters of the distribution (such as the power law exponent), 
the main purpose of our method is to distinguish between
distributions that are heavy-tailed and those that are not.
% thus in particular determining the EVT universality class of the data. 

\textit{Record statistics and record correlations.}
Given a time series $\{x_1, \dots, x_N\}$ of random
variables (RVs), the
$n^{th}$ RV is said to be a record if it exceeds all previous RVs
$\{x_j\}_{j<n}$ \cite{Glick1978,nevzorov}.  
For independent, identically distributed (i.i.d.) RVs,
it is straightforward to see that the probability $p_{n}$ for the 
$n^{th}$ entry to be a record is simply
$p_{n}=1/n$, because any of the $n$ RVs is equally likely to be
the largest. Furthermore, record events are stochastically independent in
this case \cite{Glick1978,nevzorov} and hence the \emph{joint} probability
$p_{n, n-1}$ that both $x_{n-1}$ \emph{and} $x_{n}$ are records
factorizes to $p_{n, n-1}=p_{n}p_{n-1}$ 

In a recent surge of interest
\cite{k2007,mz2008,Eliazar2009,fwk2010,Sabhapandit2011}, record
statistics has been explored beyond the classical situation of
i.i.d. RVs, and it has been found that the stochastic
independence of record events is largely restricted to the
i.i.d. case. 
In particular, for time series constructed from the linear drift 
model \cite{fwk2010,br1985}
\begin{equation}\label{ldm}
x_{n}=cn+\eta_{n}, 
\end{equation}
where $c>0$ is a constant and $\{\eta_n\}$ a family of i.i.d. RVs with
distribution $F(\eta)$ and density $f(\eta)$, correlations between
record events were quantified by considering the ratio \cite{wfk2011}
\begin{equation}
  l_{n, n-1}(c)=\frac{p_{n, n-1}(c)}{p_{n}(c)p_{n-1}(c)}.
\end{equation}
For stochastically independent record events, $l_{n,
  n-1}(c)=1$ and any positive (negative) deviation from unity can be
interpreted as the  a sign of attraction (repulsion) between record events. 
In \cite{wfk2011} both cases were found depending on the distribution
$F(\eta)$. Specifically, an expansion to first order in $c$
yields $l_{n, n-1}(c) =  1 + cJ(n) +\mathcal{O}(c^2)$ with $ J(n) \approx
-\frac{1}{2}n^4(I(n)-I(n-1))-n^3I(n) $ where 
\begin{equation}
\label{I}
I(n)=\int d\eta f^2(\eta)F^n(\eta)
\end{equation} 
and clearly $I(n) - I(n-1) < 0$. Thus for large $n$, there are two
competing contributions to $J(n)$ determining the sign of the
correlations.

To classify the behavior of the correlations in terms of the
EVT classes \cite{Sornette,dehaan}, 
consider the generalized Pareto distribution
\cite{pickands} $f(\eta)=(1+\kappa\eta)^{-(\kappa +1)/\kappa}$, which
reproduces the three classes as $\kappa<0$ (Weibull), $\kappa
>0$ (Fr\'echet) and $\kappa =0$ (Gumbel), respectively. 
Computing $I(n)$ separately for these three cases \cite{fwk2010} 
it was shown that, up to
multiplicative terms in $\log(n)$ or slower, one has $I(n)\sim
n^{-(2+\kappa)}$ and therefore \cite{wfk2011}
% \begin{equation}\label{j_fre_wei}
$J(n) \approx  \frac{\kappa}{2}n^3I(n)$,
% \end{equation} 
showing that the sign of correlations is directly determined by the extreme
value index $\kappa$ \cite{note0}. 

In the Gumbel class ($\kappa=0$) more refined calculations for 
the generalized Gaussian densities $f_{\beta}(x) \sim 
\exp(-|\eta|^{\beta})$ show that correlations are 
negative for $\beta > 1$ and positive for $\beta <1$ \cite{wfk2011}. 
The marginal case of a pure exponential distribution also shows positive
correlations, but they can be distinguished from the $\beta<1$ case in
magnitude and, more clearly, in their $n$ dependence: While for
$\beta<1$, correlations grow with $n$ up to a limiting value, for
$\beta=1$ they are independent of $n$. The special, marginal role of the
exponential distribution was also encountered in a study of near-extreme
events \cite{sm2007}, where the integral (\ref{I}) appears in a different
context.

To sum up, correlations between record events in time series
with a linear drift allow a clear distinction between underlying
probability densities that decay like an exponential or faster for
large argument, and densities with heavier tails,
by looking for positive correlations
that grow in $n$. Using these two criteria, we now present a 
distribution-free test for heavy tails in data sets of i.i.d. 
random variables. 

\textit{Description of the test.} Consider a data set
with $N$ entries, $x_1, 
x_2, \dots, x_N$ that can reasonably be argued to consist of
independent samples from the same distribution \cite{note1}. Then
for each $n<N$, one can pick uniformly at random a subset of $n$ entries 
and add a linear trend according to the index in the subset (see
Eq.(\ref{ldm})), thus forming a set of 
random variables with linear trend. 
For each $n$, there are ${N \choose n}$ possible subsets \cite{note2}, which
can be used to compute the fraction of times the $n^{th}$ entry is a
record $\hat{p}_{n}(c)$, the corresponding fraction
$\hat{p}_{n-1}(c)$ for the ${n-1}^{th}$ entry, and 
the fraction $\hat{p}_{n, n-1}(c)$ of times both entries are records, 
for each value of a suitably chosen range of $c$ \cite{note3}. The
number $s$ of subsets used for each value of $c$ will be referred to
as `internal statistics'. Finally, one obtains an estimate for the
correlations  
% \begin{equation}
  $\hat{l}_{n,
    n-1}(c)=\frac{\hat{p}_{n}(c)\hat{p}_{n-1}(c)}{\hat{p}_{n, n-1}(c)}$,
% \end{equation}
where the hat serves to indicate that we are
dealing with one fixed times series of length $N$ and its
sub-series, rather than many independent realizations. 
In the following we refer to $\hat{l}_{n,n-1}(c)$ as the 
\textit{heavy tail indicator} (HTI). 

\begin{figure}
  \includegraphics[width=0.45\textwidth]{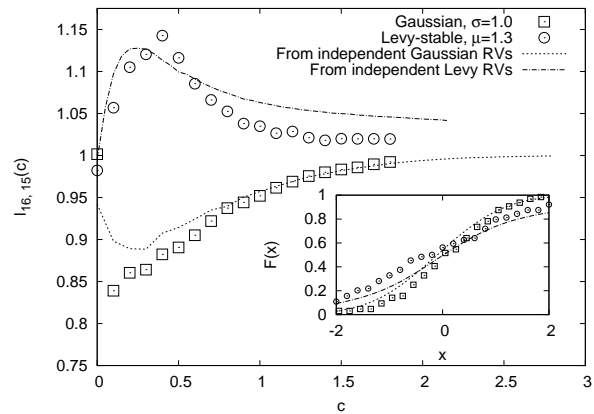}\\
  \caption{\label{example} A first example of the proposed test, with
    $N=64$ i.i.d. RVs drawn  from a Gaussian with unit variance
    (squares) and a symmetric L\'evy distribution $L_\mu(x)$ with $\mu=1.3$
    (circles). \textbf{Inset:} Comparing the cumulative distribution
    function $F(x)$ (lines) to its empirical estimate from the $64$ data points
    shows that one distribution is broader than the other but does not allow for 
a clear distinction between the two data sets. \textbf{Main plot:} This
    difference is however clearly seen under application of the
    record-based test for subsamples of size $n=16$. Dotted and dashed-dotted lines show the
    prediction for $l_{16,15}(c)$ for independent RV's.  
}
\end{figure}

To see how the test works in practice, consider Fig.~\ref{example}. 
Two data sets of size $N=64$ each are presented,
one drawn from a standard Gaussian distribution, the
other from a symmetric L\'evy stable distribution with parameter
$\mu=1.3$. A standard approach to inferring the shape of the distribution
is to estimate the cumulative distribution function by rank ordering the data
along the $x$-axis (inset). In the example this shows that one distribution is broader
than the other, but does not allow to distinguish between a difference
in scale (as for two Gaussians of different standard deviation) and a
difference in shape. In contrast, the two data sets 
come apart quite clearly under application of the test, showing
that $\hat{l}_{n, n-1}(c) > 1$ for the L\'evy distribution and 
$\hat{l}_{n, n-1}(c) < 1$ for the Gaussian (main figure).  

\begin{figure}
  \includegraphics[width=0.45\textwidth]{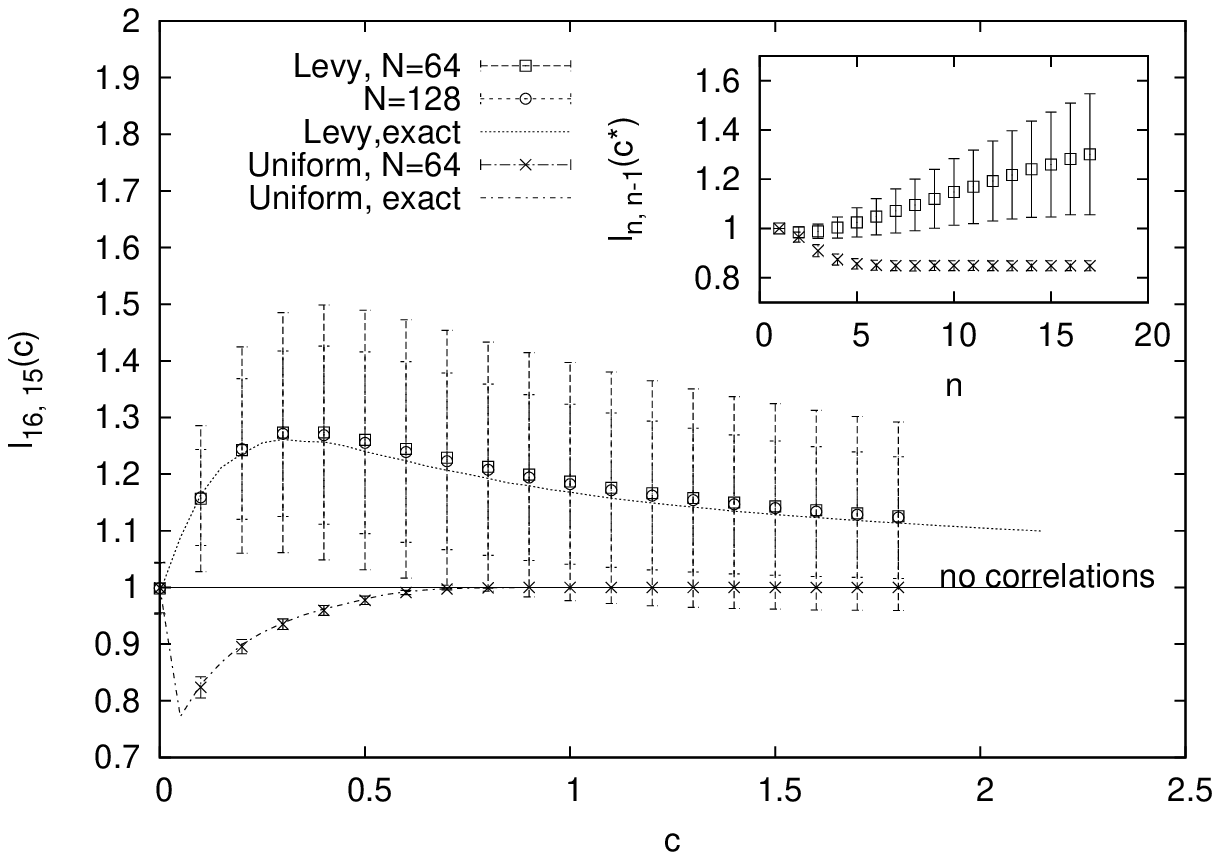}\hfill
  \includegraphics[width=0.45\textwidth]{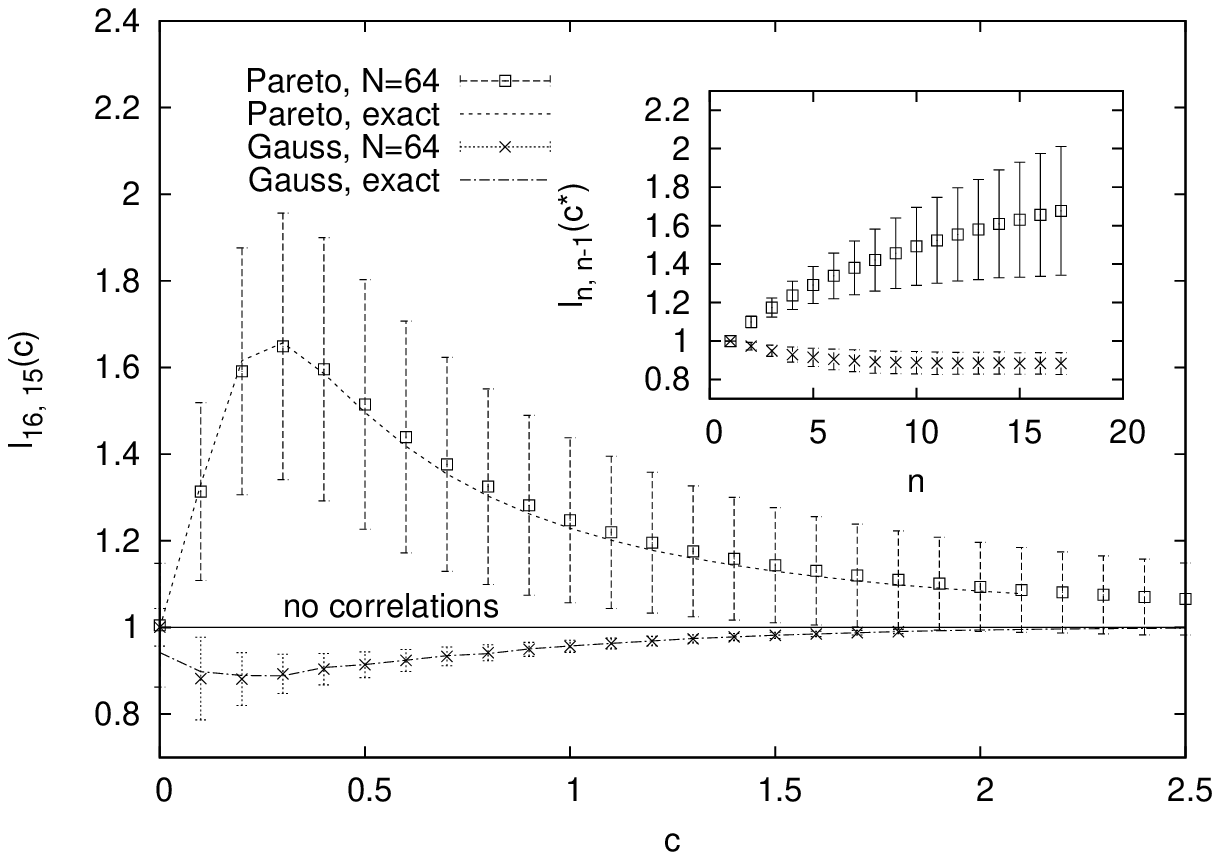}
  \caption{\label{corr_var} Sample-to-sample fluctuations of the HTI
$\hat{l}_{n,n-1}$ for different distributions. 
Lines show mean values of the correlation $l_{16,15}(c)$ obtained from simulations
of independent RV's (labeled \emph{exact}), symbols show the mean value of the HTI and error
bars indicate the standard deviation of the fluctuations for the
symmetric L\'evy-stable distribution with tail-parameter $\mu=1.3$ and
uniform distribution on $(0,1)$ (top),  and the Pareto-distribution
with $\mu=2.0$ and standard normal distribution (bottom).
The HTI was obtained  
    from simulations with internal statistics $s=10^4$
    (Pareto) or $s=10^5$ (all other) and
    averaged over $S=10^3$ independent data sets. 
    Insets show how the
    correlations at the value $c^\ast=0.25$ where correlations deviate
    maximally from unity grow as function of $n$ while keeping $N$ fixed.}
\end{figure}

\begin{figure}   
 \includegraphics[width=0.45\textwidth]{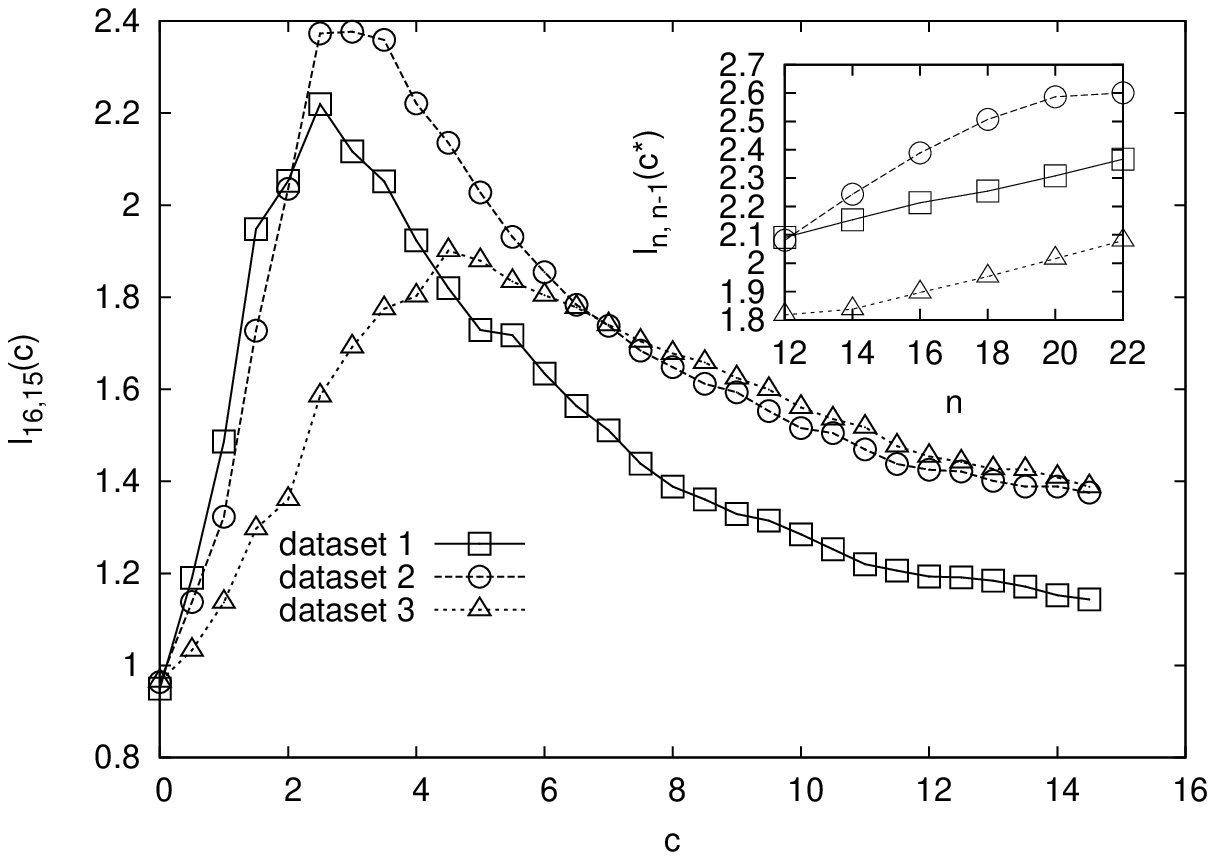}\hfill
\includegraphics[width=0.45\textwidth]{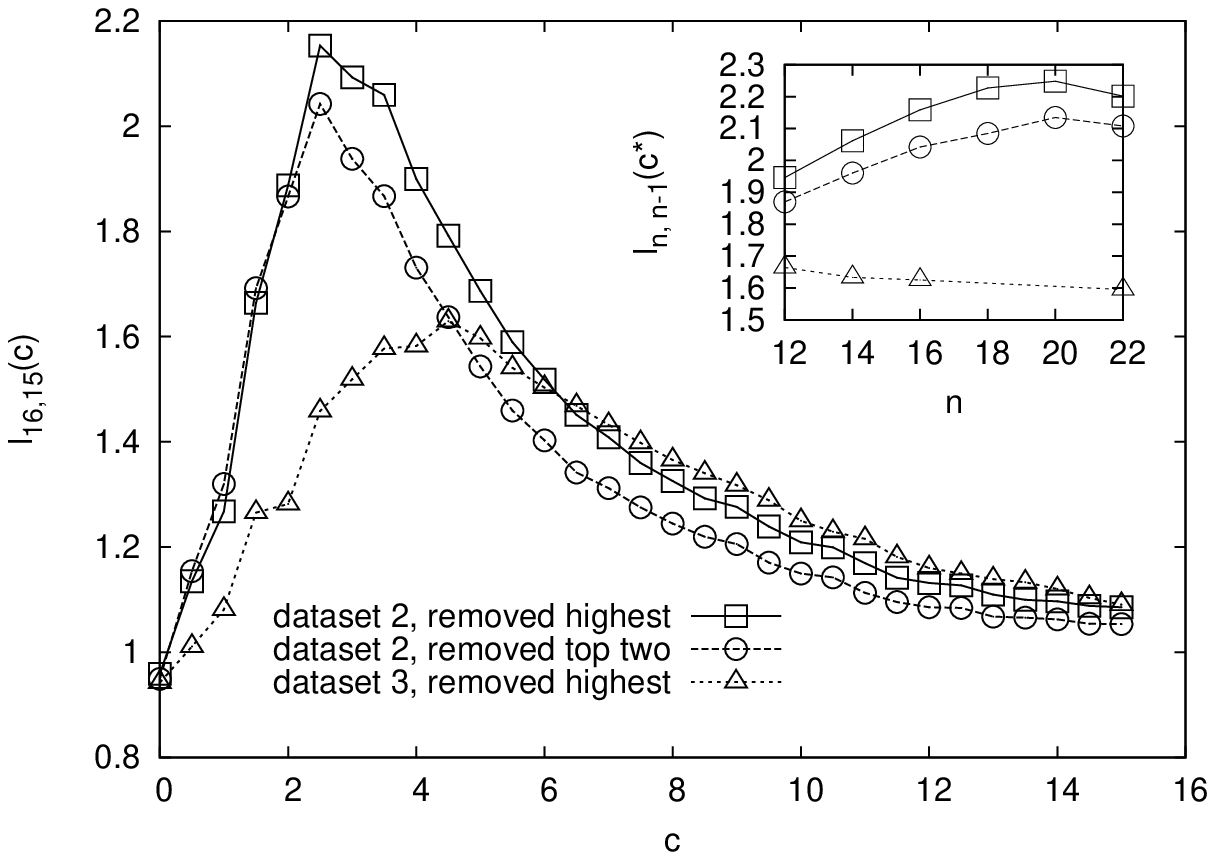}
\caption{\label{isi} \textbf{Top:} Three randomly chosen subsets of length
  $N=64$ each from the ISI citation data set \cite{r1998}. The HTI was
  computed with internal statistics $s=10^6$ and $n=16$. The main plot shows
  attractive correlations in all three cases, the inset verifies
  growth of these correlations with $n$. \textbf{Bottom:} Removing the
  largest and even the top two entries of data set 2 does not change the
  result of the test. In data set 3, which is a somewhat
  extreme case in that the largest value is more than a factor 10
  greater than the second largest, the correlations remain attractive upon removal of the largest entry 
  but the magnitude of correlations no longer increases with $n$.} 
\end{figure}

\textit{Fluctuations.}
The lines in the main part of Fig.~\ref{example} show the predicted correlation
$l_{n,n-1}(c)$ obtained from simulations of independent RV's.
The estimated HTI $\hat{l}_{n,n-1}(c)$ obtained from subsamples of the
two finite data sets deviates from these predictions, reflecting the
fact that the  ensemble of subsamples is \textit{not} independent. The
deviations depend on the data set in a random way, compare to
Fig.~\ref{isi}, and understanding how the magnitude of the deviations
depends on the test parameters $N$, $n$ and $s$ is clearly 
important for a quantitative assessment of the significance of the test. 
Figure \ref{corr_var} explores these sample-to-sample 
fluctuations by computing $\hat{l}_{n, n-1}(c)$ for a large number $S$
(`external statistics') of different data sets and recording the mean
and the mean squared deviation for different
distributions. 
The fluctuations are large for power law distributions
and decrease significantly for representatives of the Gumbel and
Weibull classes. The latter implies that it is very unlikely for positive correlations to be produced 
by chance if the underlying distribution is \textit{not} of heavy tail type; the observation
of a HTI exceeding unity can therefore be taken as a strong indication of heavy tailed behavior
in the data.

\begin{figure}
  \includegraphics[width=0.45\textwidth]{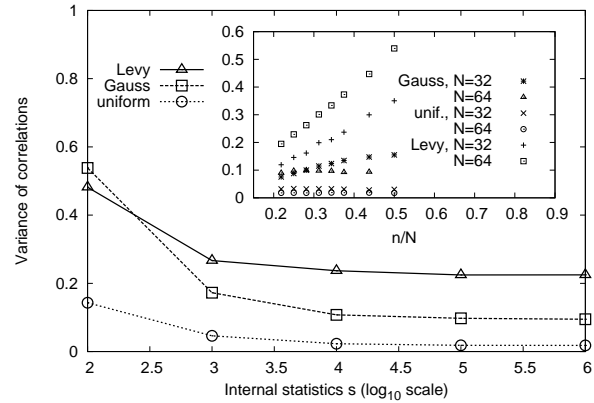}
  \caption{\label{var_conv} Magnitude of sample-to-sample fluctuations for three of the
cases considered in Fig.\ref{corr_var}. With increasing internal statistics $s$, the
    sample-to-sample fluctuations decrease to a limiting value (main
    plot). This limiting value increases with $n/N$ (inset),
    indicating that best results in terms of fluctuations are
    obtained by considering short subsequences. In the main plot
    $N=64$ and $n=16$.} 
\end{figure}

The effect of the internal statistics on the sample-to-sample 
fluctuations is quantified in Fig.\ref{var_conv}, where
their magnitude can be seen to saturate to a limiting value with increasing $s$. 
Furthermore the limiting value depends on the ratio $n/N$: The smaller a subset of the
initial data set is used, the more precise the results can be made by
using large internal statistics. This behavior underlines a particular
strength of our approach, namely that the combinatorially large
number of subsequences can be used (up to a point) to reduce fluctuations due to the
finite size of the data set. On the other hand, $n$ should not be
chosen too small, 
as the amplitude of correlations generally increases with $n$
\cite{wfk2011} (see inset of Fig.\ref{corr_var}). 
For the examples presented here, we found $n/N=1/4$ at $N=64$ to yield
the best compromise between these two contradicting requirements, see
also Fig.\ref{var_conv}.

\textit{Application.}
As an application of our approach, we consider the ISI citation
data set first analyzed by Redner \cite{r1998}, consisting of 
citation data for $783 339$ papers published in 1981 and cited 
between 1981 and June 1997. 
Due to the large size of this data set, the existence of a power law tail
with exponent $\mu \approx 2$ is well established
\cite{r1998,Clauset2009,note4}.  
Using our record-based approach, the heavy-tailed property could be recovered
by considering small, randomly chosen subsets of only $N=64$ papers
each  (Fig. \ref{isi}).  
Despite the substantial fluctuations between the three subsets, the
HTI lies clearly above unity in all cases.
The small size of the chosen subsets implies that only a few (if any)
data points in the subsets come from the extreme tails of the distribution.
The lower panel in Fig.\ref{isi} illustrates the robustness of the test with
respect to the removal of putative outliers.

\textit{Summary.}
In conclusion, in this Letter we propose a record-based 
distribution-free test for heavy tails that works particularly
well for small data sets. It was shown that the test is very
versatile and quite robust to the removal of outliers, thus complementing
standard methods like maximum likelihood estimates \cite{Clauset2009}.
While record
statistics has a long history of yielding distribution free tests
\cite{Foster1954,Glick1978,Gulati2003},  
our approach is conceptually novel in that  
we make systematic use of the combinatorial proliferation of
subsets of the original data set, which are then  
manipulated by adding a linear drift.
We expect our method to be particularly useful in situations where the
size of the data set is intrinsically limited, as in the assignement
of an EVT universality class to the distribution of beneficial
mutations in population genetics \cite{Rokyta2008,Miller2011}. In particular,
the test can be used to strengthen the evidence in favor of
heavy-tailed behavior in situations where conventional parametric 
tests have insufficient statistical power. By combining our test with 
standard approaches such as the maximum likelihood method, 
the tail parameters can then also be estimated. 

\begin{acknowledgments}

The authors would like to thank Sid Redner
for supplying the ISI data and 
Ivan Szendro for useful discussions. We acknowledge financial
support from Studienstifung des deutschen Volkes (JF), 
Friedrich-Ebert-Stiftung (GW) and DFG within BCGS and SFB 680.
\end{acknowledgments}


\begin{thebibliography}{10}
\bibitem{note0} For $\kappa>1$, corresponding to distributions without
  a mean, the correction term $J(n)$ is a decreasing function of
  $n$ and the correlations vanish asymptotically for large $n$. However for
  finite $n$ there are positive correlations of substantial magnitude,
  and the proposed test is still applicable.
\bibitem{note2} In addition permutations of the subsets can be
  considered, which are not equivalent in the presence of drift.
\bibitem{note3} The range of $c$ should be adjusted according to the typical
  spacing between entries of the time series.
\bibitem{note4} But note that the power law is limited to the extreme
  tails of the distribution containing less than 500 papers. In the
  intermediate regime, the data are better represented by a stretched
  exponential \cite{r1998}.
\end{thebibliography}
\end{document}